\documentclass[preprint]{aastex}

\newcommand{\suzaku}{\textsl{Suzaku}}
\newcommand{\asca}{\textsl{ASCA}}
\newcommand{\chandra}{\textsl{Chandra}}
\newcommand{\xmm}{\textsl{XMM-Newton}}
\newcommand{\rosat}{\textsl{ROSAT}}

\newcommand{\tycho}{\textsl{Tycho}}

\slugcomment{To appear in The ApJ Letters}

\shorttitle{Doppler-Broadened Iron Lines from Tycho's SNR}

\shortauthors{Furuzawa et al.}

\begin{document}

\title{Doppler-Broadened Iron X-ray Lines from Tycho's Supernova Remnant}

\author{Akihiro Furuzawa\altaffilmark{1}, Daisuke Ueno\altaffilmark{1}, 
  Asami Hayato\altaffilmark{2,3}, Midori Ozawa\altaffilmark{4}, 
  Toru Tamagawa\altaffilmark{2}, Aya Bamba\altaffilmark{5}, 
  John P. Hughes\altaffilmark{6}, Hideyo Kunieda\altaffilmark{1}, Kazuo Makishima\altaffilmark{2,7}, 
  Stephen S. Holt\altaffilmark{8}, Una Hwang\altaffilmark{9}, 
  Kenzo Kinugasa\altaffilmark{10}, Robert Petre\altaffilmark{9}, 
  Keisuke Tamura\altaffilmark{5},   Hiroshi Tsunemi\altaffilmark{11}, \and Shigeo Yamauchi\altaffilmark{12} 
}

\email{furuzawa@u.phys.nagoya-u.ac.jp}

\altaffiltext{1}{Division of Particle and Astrophysical Science,
  Graduate School of Science, Nagoya University, Furo-cho, Nagoya
  464-8602, Japan}
\altaffiltext{2}{RIKEN, 2-1 Hirosawa, Wako, Saitama 351-0198, Japan}
\altaffiltext{3}{Department of Physics, Tokyo University of Science,
  1-3 Kagurazaka, Shinjuku-ku, Tokyo 162-8601, Japan}
\altaffiltext{4}{Department of Physics, Graduate School of Science,
  Kyoto University, Kita-Shirakawa, Sakyo-ku, Kyoto 606-8502, Japan}
\altaffiltext{5}{Institute of Space and Astronautical Science, Japan
  Aerospace Exploration Agency, 3-1-1 Yoshinodai, Sagamihara, Kanagawa
  229-8510, Japan}
\altaffiltext{6}{Department of Physics and Astronomy, Rutgers
  University, 136 Frelinghuysen Road, Piscataway, NJ 08854-8019 USA}
\altaffiltext{7}{Department of Physics, The University of Tokyo, 7-3-1
  Hongo, Bunkyo-ku, Tokyo 113-0033, Japan}
\altaffiltext{8}{F. W. Olin College of Engineering, Needham, MA
  02492, USA}
\altaffiltext{9}{NASA Goddard Space Flight Center, Greenbelt, MD
  20771, USA}
\altaffiltext{10}{Gunma Astronomical Observatory, 6860-86, Nakayama,
  Takayama-mura, Agatsuma-gun, Gunma 377-0702, Japan}
\altaffiltext{11}{Department of Earth and Space Science, Graduate
  School of Science, Osaka University, 1-1 Machikaneyama, Toyonaka,
  Osaka 560-0043, Japan}
\altaffiltext{12}{Faculty of Humanities and Social Sciences, Iwate
  University, 3-18-34 Ueda, Morioka, Iwate 020-8550, Japan}

\begin{abstract}

We use \suzaku\ observations to measure the spatial variation of the
Fe K$\alpha$ line with radius in the \tycho\ supernova remnant.
The Fe line widths show a significant decrease from a FWHM value of
210 eV at the center to 130 eV at the rim.  Over the same radial range
the line center energy remains nearly constant.  These observations
are consistent with a scenario in which the shell of Fe-emitting
ejecta in \tycho\ is expanding at speeds of 2800--3350 km s$^{-1}$.
The minimum line width we measure is still a factor of two larger than
expected from a single component plasma emission model.  If thermal
Doppler broadening is the dominant additional source of broadening, we
infer an ion temperature of $(1--3) \times 10^{10}$ K.

\end{abstract}


\section{Introduction}
\label{sec:intro}

Type Ia supernovae (SNe) play an important role in the chemical
evolution of the universe by providing a significant fraction of the
Fe group elements in stars, the interstellar medium and the
intracluster medium. They have also become a prime tool to explore the
expansion history of the universe. Yet, in spite of their importance,
the physical processes involved in the actual explosions remain
unclear.

The SN seen by \tycho\ Brahe in 1572 can be classified with some degree
of confidence as Type Ia based on the light curve and color evolution
from the historical record \citep{ruiz04}.  The remnant (hereafter
\tycho) is located at Galactic coordinates $(l, b)$ =
(120$^\circ$.0879, 1$^\circ$.4460) and its estimated distance is
1.5--3.1 kpc \citep{chevalier80,albinson86,smith91,ruiz04}, but still
debated \citep{schwarz95}.

\tycho\ is now a young and X-ray bright supernova remnant (SNR) that has
been studied extensively for investigations of the SN Ia explosion
mechanism. In the radio and X-ray bands, \tycho\ shows a relatively
smooth and regular $8^\prime$ diameter limb-brightened shell with
several protuberances and indentations notably toward the southeast
(SE) and northeast (NE) parts of the shell, possibly driven by
fingers of SN ejecta or the consequence of the blast wave interacting
with a nonuniform ambient medium. The good angular resolution and
photon statistics of \tycho\ observations by \chandra\ and \xmm\ have
enabled detailed studies of the spatial structure \citep{warren05} and
have set significant constraints on allowed explosion models by
comparing observed and simulated X-ray spectra
\citep{badenes06}. Among the results revealed by \chandra\ is one that
finds the region between the forward shock and contact discontinuity
to be very narrow and dominated by nonthermal emission, concentrated
in geometrically thin filamentary structures
\citep{hwang02,bamba05,warren05,cassam07}.  No thermal emission has
yet been detected from the forward shock region; all the observed
metal lines in the X-ray spectrum are produced predominantly by the
ejecta.

A direct measurement of the expansion velocity of \tycho\ has not yet
been obtained. \citet{reynoso97} reported the expansion rate obtained
from the radial displacement of the radio-emitting shell over the
course of a decade. \citet{hughes00} obtained the rate from a
difference of brightness profiles measured from two \rosat\ HRI
observations taken in 1990 and 1995. These studies indicate expansion
velocities of the outer rim of \tycho\ of 2000--4300 km s$^{-1}$ (radio)
and 2200--4600 km s$^{-1}$ (X-ray), where the large range is dominated
by the distance uncertainty. \citet{reynoso97} found a variation with
azimuth of the outer rim expansion rate and expansion parameter from
0.25 to 0.75. Combining a distance of 2.3 kpc with their age (416 yr)
and their radii leads to a range of observed rim velocities versus azimuth
of 1700--4300 km s$^{-1}$.

\tycho\ shows a strong Balmer line with a profile consisting of narrow
and broad components. The width of the broad component and intensity
ratio have been used to measure the shock velocity without distance
dependence. \citep{chevalier80,smith91, ghavamian01}. These studies
yield velocities of 1800--1900 km s$^{-1}$ (H$_\alpha$ width) and
1800--3500 km s$^{-1}$ for the optical bright knot "g".

In this Letter we determine, for the first time, the expansion
velocity of the Fe ejecta in \tycho\ from Doppler broadening of the Fe K
lines, utilizing the modest imaging capability and well calibrated
instrument response of the \suzaku\ CCD cameras \citep{ozawa09}. We
refer the reader to \citet{tamagawa09} for more details on the
broadband continuum emission of \tycho\ and the discovery of faint lines
from the Fe-group elements Cr and Mn in the integrated
\suzaku\ spectra.

\section{Observation}
\label{sec:obs}

\suzaku, the fifth Japanese X-ray astronomy satellite
\citep{suzaku07}, was launched into a 550 km high orbit on 2005 July
10. It contains two types of functioning instruments: X-ray CCD
cameras \citep[XIS]{xis07} covering the 0.2--12 keV energy range, and a
nonimaging hard X-ray detector \citep[HXD]{hxd07a, hxd07b} sensitive
to X-rays in the 10--70 keV band (PIN diodes) and 30--600 keV band (GSO
scintillators). The four XIS cameras are each located at the focus of
an independent X-ray telescope \citep[XRT]{xrt07}. One of the four XIS
cameras utilizes a back-illuminated (BI) CCD, whereas the other three
use front-illuminated (FI) CCDs. (N.B., one of the FI CCD
cameras mentioned was lost in 2006 November.)

\tycho\ was observed by \suzaku\ on 2006 June 27--29 during the time
allocated to the Science Working Group. The XIS was operated in normal
full-frame clocking mode without spaced-row charge injection
\citep{uchiyama09}.

We started with data processed through the standard pipeline (version
2.0.6.13) using calibration reference files  publicly released on 2008
February 1. 
Data accumulated during passage through, and for an additional 436 s
after emerging from, the South Atlantic Anomaly (SAA) were discarded
as were orbital periods of low cut-off rigidity (less than 6 GV). We
also rejected data taken at low elevation angles to the Earth's rim
(less than $20^\circ$ from the limb of the sunlit Earth and less than
$5^\circ$ from the rim of the dark Earth).  Hot and flickering pixels
were removed with SISCLEAN and only events with grades of 0, 2, 3, 4
and 6 were retained.

\suzaku\ also observed an offset region close to \tycho\ on 2006 June
29--30 for purposes of background estimation. We applied the same
screening criteria and event selection described above to this
background observation.  Two faint unidentified X-ray sources appear
in the offset pointing roughly 4' and 7' southwest of the center of
the field of view (FOV) \citep[see Figure 2 in][]{tamagawa09}. These
sources are located in the \tycho\ annular analysis regions numbered 6
and 7 (see Section \ref{sec:annular_spec} below). We can, however,
safely neglect these sources, since their emission is 10 times fainter
than that of \tycho\ in the 2--5 keV band.

The net exposures of on-source and offset data after the data
reductions are 101 ks and 51 ks, respectively. Details of these
observations have been summarized in \citet{tamagawa09}.

\section{Spatially Resolved Line Spectra} \label{sec:annular_spec}

\begin{figure}
  \epsscale{0.5}
  \plotone{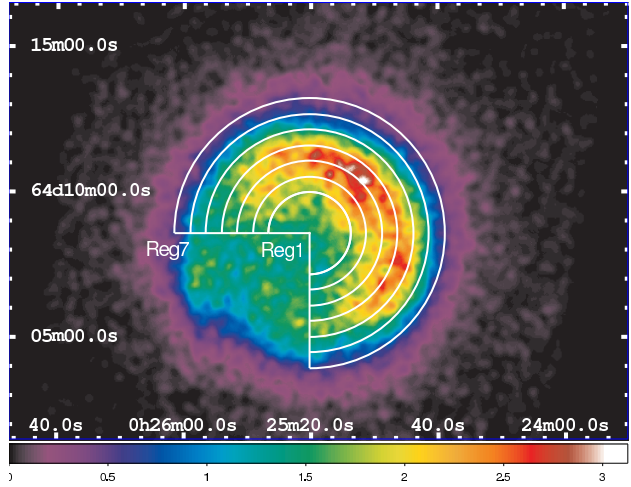}
  \caption{6--7 keV Fe-band X-ray image of the \tycho\ SNR
    obtained with the \suzaku\ XIS.  Regions defined for spectral
    analysis (Section \ref{sec:annular_spec}) are indicated.
    \label{fig:xis_image_all}
  }
\end{figure}

Guided by the nearly circular, shell-like structure of \tycho\, we
divided the XIS FOV into a central region 1'.41 in radius, surrounded
by six additional annular regions each 0'.54 wide in order to examine
radial variations in the spectral parameters. Since the structure of
the remnant is distorted toward the SE, as clearly seen in the
\chandra\ and \xmm\ observations \citep{decourchelle01,warren05}, we
have excluded the SE quadrant, as indicated in Figure
\ref{fig:xis_image_all}. The remaining NE, NW, and SW quadrants of
\tycho\ show a similar shell-like morphology in the Fe K line
\citep{warren05}.  Background spectra were accumulated from the offset
observation using the same region in detector space.

Our spectral model consists of a power-law component \citep[which
  which well approximates the continuum emission in \tycho\,
  see][]{tamagawa09} and several Gaussians to account for the
prominent emission lines.  At the energy resolution of the XIS, most
observed line features include multiple emission lines. In order to
reduce the effects of line blending from different transitions or even
different elemental species, the spectral fits were performed in the
following narrow bands separately: 1.784--1.894 keV for He-like Si
K$\alpha$, 2.40--2.60 keV for He-like S K$\alpha$ and 6.00--6.90 keV
for Fe K$\alpha$. For this Letter, we focus primarily on the Fe K line
profiles.  The lower energy lines, and Si He-like K$\alpha$ in particular,
are subject to complexities in the calibration of the energy scale,
and thus require a much more careful analysis.  We defer this to a
subsequent paper, but use the Si and S line profiles for illustrative
purposes only in this Letter.

Best-fit values and uncertainties (at 90\% confidence) are quoted in
Table \ref{tab:fit} for the Fe K$\alpha$ spectra (labeled from the
center, Reg1, to the rim, Reg7) and Figure \ref{fig:raddep_gauss_all}
shows the radial dependence of these parameters.  The average centroid
energy we find, $6455$ eV is consistent with previous measurements by
\asca\ and \xmm\ \citep{hwang97,badenes06}. \citet{tamagawa09} used
the same \suzaku\ observation data and obtained the centroid over the
whole remnant as $6445$ eV. The difference of 10 eV is caused by using
different calibration reference files, since calibration reference
files publicly released in 2007 were used in \citet{tamagawa09}.
The line centroids prefer a slight radial gradient corresponding to an
increase of $\sim$12 eV from the center to the rim.  The line widths on the
other hand show a highly significant gradient from a FWHM value of 210
eV near the center to values of $\sim$140 eV at the rim. A flat
profile can be rejected with high confidence (greater than $3 \sigma$).

To estimate the calibration error of the current response function,
the width of the Mn K$\alpha$ calibration line from the built-in
calibration source ($^{55}$Fe), which illuminates the two far-end
corners from the readout node, has been examined. The width of the
calibration line has been obtained to be $16 ^{+12}_{-16}$ eV FWHM.
In the source detected region, the broadening due to the charge
transfer inefficiency (CTI) is smaller than that of the calibration
source because of the reduced number of charge transfers. Our
obtained widths (130--210 eV) in \tycho\ are significantly larger than
that of CTI induced broadening.

We note that because of the broad point-spread function (PSF) of the
\suzaku\ XRT (half-power diameter of 2$^\prime$) the radial bins used
here are not fully independent.  To assess the effects of smearing,
we used XISSIM to simulate an XIS image assuming a thin shell model
for the radial surface brightness profile of the Fe K$\alpha$ line
intensity \cite[as given in][]{warren05}.  Only 25\% of the observed
events in region 6 actually originate from this sky region; 60\% come
from regions 4 and 5 where the Fe K$\alpha$ brightness
peaks. Virtually all of the photons in region 7 (the rim) come from
regions 4, 5 and 6.  Thus any true intrinsic spatial variation of the
line parameters in the outermost three radial bins has been smeared
out by the XRT PSF.  On the other hand, smearing effects of the XRT
are much less important for the variation interior to radii of
$\sim$3$^\prime$.

The ionization age can be estimated from the line center energy. Our
measured central energies of 6448--6461 eV and widths of 130--210 eV
FWHM indicate the existence of low ionized Fe below
Fe$_\mathrm{XVIII}$.  Of the nonequilibrium plasma emission models in
XSPEC, only the NEI version 1.1 model includes low ionized (below
He-like) ions. We therefore use this model. Figure \ref{fig:nt_e_fwhm}
shows the ionization age dependence of the centroid energy and width
at various electron temperatures when the modeled Fe K$\alpha$ blend
is fitted by a single Gaussian line.

Our measured central energies of 6448--6461 eV correspond to Fe
ionization ages in the range $10^{10} < n_\mathrm{e} t
/(\mathrm{cm^{-3}~s)} < 10^{11}$ for all electron temperature values
$k T_\mathrm{e} > 0.6$ keV. Simulated spectra over this range of
thermodynamic parameters yield expected Gaussian widths for the Fe
K$\alpha$ blend of only 50--110 eV FWHM. These values are far below
what we measure in the central region ($\sim$ 210 eV) and even at the
rim ($\geq $130 eV), implying the presence of additional line broadening
mechanisms in \tycho\.  Results from the independent Si and S band
analyses also show similar tendencies in the radial dependence of line
centroids and widths.  Calibration uncertainties and line blending
complicate the analysis of these lower energy lines; we defer a
quantitative discussion of these results to a future study.

In order to directly compare the line profiles without model fitting,
we took the ratio of the narrowband spectrum in region 7 (rim) to
that in region 1 (center) for the Si, S and Fe K$\alpha$ lines
(Figure \ref{fig:line_profile_ratio}). All three ratios show a similar
profile characterized by an ``M'' shape (lower in the middle, higher
at the edges).  The ratios show that the line shape in the central
region is significantly broader than at the rim for all three lines.

To investigate further the broad Fe K$\alpha$ line profile, we next
fitted a double Gaussian line plus power-law continuum model to the
spectrum extracted from the remnant's center.  Both Gaussian widths
were fixed at the value obtained from the rim (130 eV FWHM). The fits
were acceptable as shown in Figure \ref{fig:double_gauss_Fe} with
best-fit center energy values of $6405^{+15}_{-14}$ eV and
$6536^{+27}_{-19}$ eV and normalizations of $3.1^{+0.7}_{-0.5} \times
10^{-4}$ photons cm$^{-2}$ s$^{-1}$ and $2.1^{+0.6}_{-0.7} \times
10^{-4}$ photons cm$^{-2}$ s$^{-1}$, respectively. The difference in
line energies corresponds to a velocity difference of $\pm
3040^{+310}_{-240}$ km s$^{-1}$ or it can be interpreted as
contributions from ionization timescales of $n_\mathrm{e} t \approx 1
\times 10^9$ cm$^{-3}$ s and $n_\mathrm{e} t \approx (5--10) \times
10^{10}$ cm$^{-3}$ s. When we change the fixed width within the 90\%
confidence level (100--150 eV FWHM), the center energy values vary by
only $\pm 7$ eV, well within the 90\% error range.

\section{Discussion and Conclusion}

The excellent energy resolution of the \suzaku\ XIS has enabled us to
discover significant broadening of the Si, S and Fe emission lines in
\tycho\, as well as a clear decrease in the line widths from the center
toward the rim.  In the following discussion we consider, in turn, two
simple astrophysical interpretations of these observations: (1) radial
variations of the thermodynamic state of the X-ray-emitting plasma and
(2) expansion of a spherical shell.

\begin{figure}
  \epsscale{0.5}
  \plotone{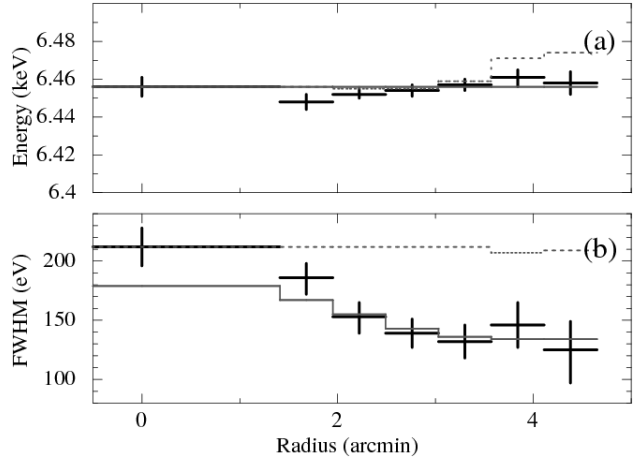}
  \caption{Radial dependence of best-fit line energies (a) and widths
    (b). Data points were obtained from spectra extracted from the sum
    of the NE, NW, and SW segments indicated in
    Figure~\ref{fig:xis_image_all}. Solid and dashed gray lines
    represent the roughly estimated profiles of the expanding shell
    model and the two-zone model, respectively.
    \label{fig:raddep_gauss_all}
  }
\end{figure}

\begin{figure}
  \epsscale{0.5}
  \plotone{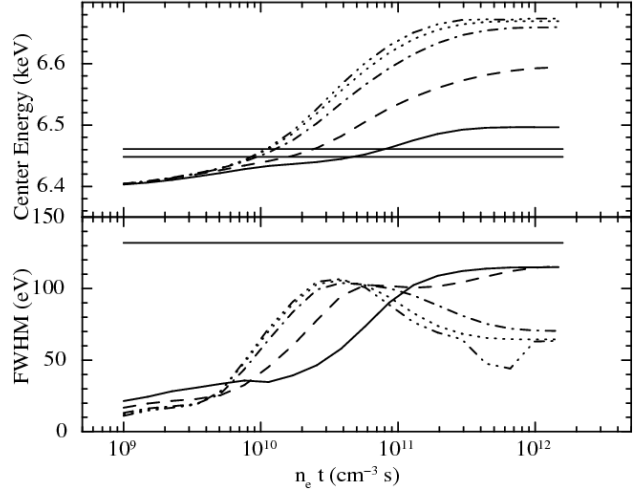}
  \caption{Fitted Gaussian center energies and widths as a function of
    ionization timescale obtained from simulated nonequilibrium
    ionization plasma spectra with NEI model (version 1.1) at the
    electron temperatures of 0.6 (solid), 1.0 (dashed), 2.0
    (dashed-dotted), 3.0 (dotted), and 4.0 keV
    (dashed-dotted-dotted-dotted).  Horizontal lines in the upper
    panel show the range of observed center energies. The line in the
    bottom panel shows the measured FWHM of Region 5 for reference.
    \label{fig:nt_e_fwhm}
  }
\end{figure}

\begin{figure}
  \epsscale{0.5}
  \plotone{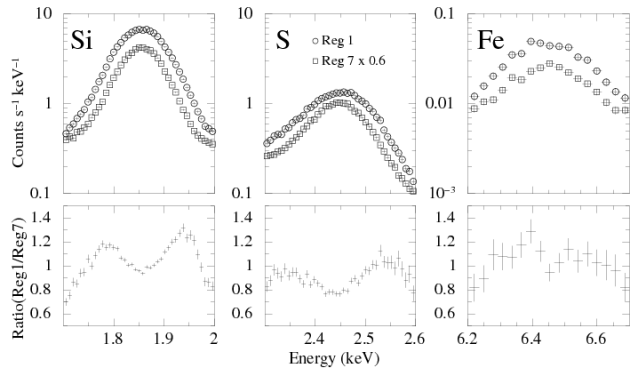}
  \caption{Spectra and spectral ratios for bands containing the Si, S
    and Fe K$\alpha$ lines. Top: spectra in regions 1 (open circles)
    and 7 (open squares). Spectra in region 7 were multiplied by
    0.6. Bottom: ratios of spectra in region 1 to those in region 7.
    \label{fig:line_profile_ratio}
  }
\end{figure}

\begin{figure}
  \epsscale{0.5}
  \plotone{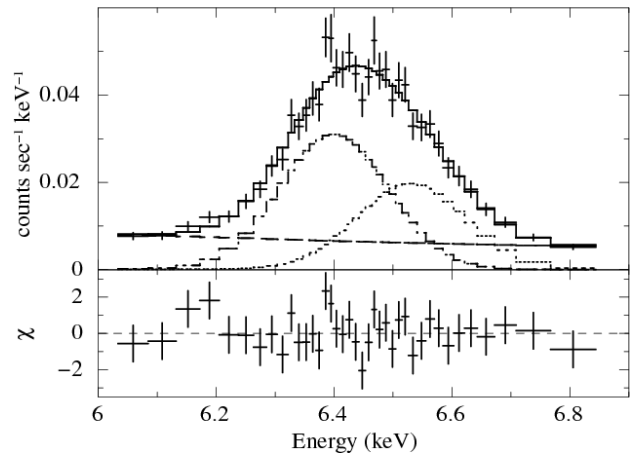}
  \caption{Spectrum covering the 6.0--6.9 keV band with 
    the best-fit double Gaussian and power-law model. Residuals 
    (bottom panel) are in units of $\chi$.
    \label{fig:double_gauss_Fe}
  }
\end{figure}

The presence of two Gaussian lines in the central region can be
explained naturally by the presence of two (or more) different plasma
states: hot gas with low ($n_\mathrm{e} t \approx 1 \times 10^9$
cm$^{-3}$ s) and high ($n_\mathrm{e} t \approx (5--10) \times 10^{10}$
cm$^{-3}$ s) ionization timescales.  There is some independent
evidence for this, resting on the observation that the Fe K$\alpha$
emission in \tycho\ peaks interior to that of Fe L and Si K$\alpha$
\citep{hwang97,decourchelle01,hwang02}.  The theoretical
interpretation \citep{badenes06} is that the shocked ejecta exhibit a
radial temperature and ionization timescale variation in the sense
that near the contact discontinuity the ejecta are cool with high
ionization age, while near the reverse shock they are hot with low
ionization age.  Guided by this theoretical picture, let us consider a
simple two-zone model in which the interior half of the Fe-emitting
shell produces a 6.41 keV line while the exterior half produces a line
at 6.54 keV.  The relative proportions of the emission from each of
these shells seen in projection will vary from the center to the rim,
which will produce a radial change in the observed line width.
However our calculations show that the broad PSF of the telescope
smears out the radial variation in width which, under this scenario,
occurs most rapidly near the rim where the differences in the
projected surface brightnesses of the emission zones are most
extreme. This is unlike the smooth gradual drop with radius in the
observed profile.  Another problem for this model is that it produces
a fairly large shift (greater than 30 eV) in the centroid of the Fe
K$\alpha$ line between the center and the rim. This is somewhat larger
than the shallow gradient of $\sim$12 eV observed in \tycho\. In
Figure~\ref{fig:raddep_gauss_all}, profiles expected for the two-zone
model (with assumptions of low ionized Fe shell (6.41 keV line) of
3$^\prime$--3$^\prime$.5 and high ionized shell (6.54 keV line) of
3$^\prime$.5--4$^\prime$) are shown. It suggests that the two-zone
model cannot describe the constant center energy in the outer portion
and gradual drop of the width.

If the Fe emission comes from a radially expanding shell, then the
projection of redshifted and blueshifted lines from the rear
(receding) and front (approaching) parts of the shell will produce a
double line near the center.  Furthermore, the energy separation of
the lines depends on the velocity component along the line of sight,
which varies as a cosine function with projected distance from the
center.  At the rim the shell velocity is transverse to the
line of sight and so the line width there should be minimum.  In the
ideal case the energy centroid of the line should be constant with
radius, although deviations from spherical symmetry can introduce some
variation.  In general the centroid and width profiles of the Fe
K$\alpha$ line we measure in \tycho\ are more consistent with an
expanding shell interpretation than multiple thermodynamic emission
zones. In Figure~\ref{fig:raddep_gauss_all}, the radial profiles
expected for the expanding shell model (with assumptions of expansion
velocity of 3000 km s$^{-1}$ and radius of the shell of $3^\prime$)
are shown.

The minimum line width of 130 eV FWHM we obtain at the rim, where
velocity broadening should be minimal, is still wider than that
expected from a single-component NEI model (50--110 eV FWHM).  An
additional source of broadening, at the level of 70--120 eV, is
therefore required. Under the assumption that this comes fully from
thermal Doppler broadening (i.e., ignoring macroscopic turbulence), we
estimate the ion temperature to be $k T_\mathrm{ion} = $ (1--3) $\times
10^{10}$ K.  This is consistent with the shock temperature expected in
models that well describe the ejecta spectrum of \tycho\ \citep[see
  Figure~8 in][]{badenes06}.

Our observations thus indicate that the Fe-emitting ejecta of the
\tycho\ SNR is expanding at a velocity of $\sim$3000 km s$^{-1}$. If
macroscopic mixing of the ejecta has not taken place, then we would
expect the intermediate mass elements (Si and S) to show even larger
velocities. Estimating the expansion speed for these other species is
going to be important for further investigation of the
three-dimensional structure and dynamics of \tycho\. For that purpose,
careful calibration of the energy scale around the Si K$\alpha$
emission line and more precise modeling of the X-ray spectrum are
needed. More detailed results including these other elements will be
presented in a forthcoming paper (A. Hayato et al. 2009, in
preparation).

\acknowledgments

We thank all members of the \suzaku\ team for their careful work
operating the satellite and calibrating the instruments.  J.P.H.
acknowledges support from NASA grant NNG05GP87G. A.H. and M.O. are
financially supported by the Japan Society for the Promotion of
Science.

\newpage

\begin{deluxetable}{ccc}
  \tabletypesize{\scriptsize}
    \tablecaption{Best-fit single Gaussian fits to \tycho\ Fe K$\alpha$ line\label{tab:fit}}
\tablehead{
  \colhead{Region} & 
  \colhead{Centroid} & 
  \colhead{Width (FWHM)}  \\
  \colhead{} & 
  \colhead{(eV)} & 
  \colhead{(eV)} 
}
\startdata
Reg1      & $6456^{+5}_{-5}$ & $212^{+16}_{-16}$ \\
Reg2      & $6448^{+4}_{-4}$ & $186^{+12}_{-14}$ \\
Reg3      & $6452^{+3}_{-2}$ & $153^{+12}_{-14}$ \\
Reg4      & $6454^{+3}_{-3}$ & $139^{+12}_{-12}$ \\
Reg5      & $6457^{+3}_{-3}$ & $132^{+14}_{-14}$ \\
Reg6      & $6461^{+4}_{-5}$ & $146^{+19}_{-19}$ \\
Reg7      & $6458^{+6}_{-6}$ & $125^{+24}_{-28}$ \\
\enddata
\tablecomments{All errors are quoted at the 90\% confidence level.}
\end{deluxetable}

\clearpage

\end{document}